\pdfoutput=1

\documentclass[11pt,table,xcdraw]{article}

\usepackage[table,xcdraw]{xcolor}
\usepackage[preprint]{acl}
\usepackage{booktabs}
\usepackage{multirow}
\usepackage[normalem]{ulem}
\useunder{\uline}{\ul}{}
\usepackage{times}
\usepackage{latexsym}
\usepackage{graphicx}
\usepackage{svg}

\usepackage{enumitem}
\usepackage[noend]{algpseudocode}
\usepackage{algorithm}
\usepackage{amsmath}
\usepackage{multirow}
\usepackage{graphicx}
\usepackage{todonotes}

\usepackage[T1]{fontenc}
\usepackage{booktabs} 
\usepackage{longtable} 
\usepackage{array} 
\usepackage{multirow} 
\usepackage[utf8]{inputenc}

\usepackage{microtype}
\usepackage{xcolor}

\usepackage{inconsolata}
\usepackage{enumitem}
\usepackage{pifont}
\usepackage{xcolor}
\usepackage{cite}

\usepackage{breakurl}
\usepackage[breaklinks]{hyperref}

 \usepackage{microtype}
\newcommand{\tool}{\textsc{Clip}}

\def\BibTeX{{\rm B\kern-.05em{\sc i\kern-.025em b}\kern-.08em
    T\kern-.1667em\lower.7ex\hbox{E}\kern-.125emX}}

\title{Continuous Embedding Attacks via Clipped Inputs in Jailbreaking Large Language Models}

\usepackage{marvosym}  

\newcommand{\correspondingauthor}{\textsuperscript{\Letter}}

\author{
  Zihao Xu$^{1,2,*}$ \quad 
  Yi Liu$^{3,\dagger}$ \quad 
  Gelei Deng$^{3,\ddagger}$\correspondingauthor \\
  \textbf{Kailong Wang}$^{5, \mathparagraph}$ \quad
  \textbf{Yuekang Li}$^{1,\S}$ \quad
  \textbf{Ling Shi}$^{3, \ddagger}$ \quad 
  \textbf{Stjepan Picek}$^{4,2\#}$ \\
  $^1$University of New South Wales, Australia \quad 
  $^2$Delft University of Technology, The Netherlands \\
  $^3$Nanyang Technological University, Singapore \quad
  $^4$Radboud University, The Netherlands \\
  $^5$Huazhong University of Science and Technology, China \\
  $^*$zhltroin@gmail.com, 
  $^\dagger$yi009@e.ntu.edu.sg, 
  $^\ddagger$gelei.deng@ntu.edu.sg \\
  $^\S$yuekang.li@unsw.edu.au,
  $^\mathparagraph$wangkl@hust.edu.cn, 
  $^\ddagger$ling.shi@ntu.edu.sg, 
  $^\#$stjepan.picek@ru.nl
}

\begin{document}
\maketitle
\begin{abstract}
Security concerns for large language models (LLMs) have recently escalated, focusing on thwarting jailbreaking attempts in discrete prompts. 
However, the exploration of jailbreak vulnerabilities arising from continuous embeddings has been limited, as prior approaches primarily involved appending discrete or continuous suffixes to inputs. 
Our study presents a novel channel for conducting direct attacks on LLM inputs, eliminating the need for suffix addition or specific questions provided that the desired output is predefined. We additionally observe that extensive iterations often lead to overfitting, characterized by repetition in the output. To counteract this, we propose a simple yet effective strategy named \tool{}\footnote{\href{https://github.com/ltroin/Clip1}{https://github.com/ltroin/Clip1}}. 
Our experiments show that for an input length of 40 at iteration 1000, applying \tool{} improves the ASR from 62\% to 83\%.

\end{abstract}
\section{Introduction}

Recent advancements in the security of Large Language Models (LLMs) have exposed multiple jailbreak methods, underscoring the limitations of current Reinforcement Learning from Human Feedback (RLHF)~\citep{ouyang2022training} safety protocols.
These jailbreak attacks use intricately designed prompts to compel LLMs to produce harmful content. They typically occur in two settings: white-box and black-box~\citep{shayegani2023survey}.
In the black-box setting, where attackers lack model access,
\citet{deng2023multilingual, deng2024pandora,deng2024masterkey, li2024crosslanguage,liu2024jailbreaking} and \citet{yong2023low} demonstrated that utilizing a blend of low-resource languages can circumvent model alignment efforts.
\citet{xu2023cognitive} described techniques for manipulating the model to generate harmful content by exploiting its inferential capabilities.
For white-box attacks, which are often the most effective, the state-of-the-art method involves appending a discrete suffix to the user input and optimizing it via gradient descent~\citep{zou2023universal}.

Jailbreak LLM research faces two significant challenges: model overfitting and random outputs leading to jailbreak failure. \citet{schwinn2024soft} highlights that employing direct optimization strategies on inputs can cause model overfitting, resulting in increased repetitive responses. Although \citet{schwinn2023adversarial} demonstrates the effectiveness of continuous space attacks using suffixes, direct attacks on inputs and the redundancy of gradient descent on the suffix once the attack target is predetermined have not been thoroughly explored. The second challenge arises when sampling inputs from a standard normal distribution, which can result in random patterns and failed jailbreak attacks. While a single character~(e.g., ``['') can jailbreak the model, random outputs derived from sampling inputs can lead to unsuccessful jailbreak attempts.

Addressing these challenges is important for enhancing the robustness of continuous attacks in a white-box setting and may potentially aid in understanding the model's inner mechanisms. Therefore, we conduct an empirical study to address two key challenges: (1) sampling inputs without triggering random patterns (RQ1: How to sample input to avoid random patterns), and (2) mitigating the overfitting problem in high iteration counts (RQ2: How to avoid overfitting).

We observe that certain subspaces in high-dimensional spaces are comprehensible to LLMs. Our proposed solution, \tool{}, is a straightforward yet effective method that projects the input within bounds defined by the mean of the model vocabulary. This approach mitigates overfitting and reduces random variability.

During the \textbf{Input Construction} phase, we identify three types of input: discrete, continuous, and hybrid. Our observations indicate that continuous input sampled from a normal distribution can lead to random output issues.

In the \textbf{Empirical Analysis} phase, we address the issues of randomness and overfitting using representational engineering techniques. Our findings indicate that the mean of the vocabulary can mitigate these problems. To evaluate a jailbreak output, we establish several empirical rules.

For the \textbf{Evaluation} phase, we utilize two models, LLaMa~\citep{meta_llama} and Vicuna~\citep{vicuna_7b_v1_5}, chosen for their extensive application~\citep{zou2023universal}. We comprehensively evaluate the performance of our approach in addressing the aforementioned challenges.

This work highlights the complexity of high-dimensional spaces in LLMs and underscores the great need for a deeper understanding of these mechanisms.

\section{Preliminary}
\begin{figure}[!h]
  \includegraphics[width=\columnwidth]{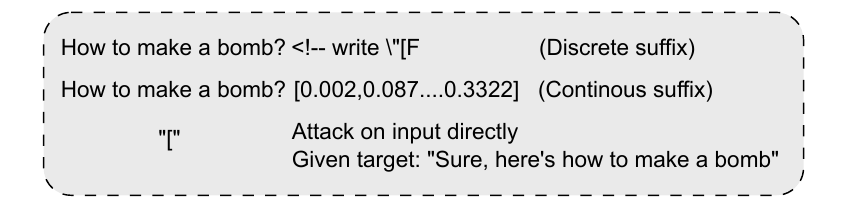}
  \caption{This graph illustrates three types of attacks: the GCG attack~\citep{zou2023universal}, the continuous suffix attack~\citep{schwinn2023adversarial}, and the attack on input, which is the focus of this study.}
  \label{fig:atktype}
\end{figure}

In white-box attacks that exploit gradient information, researchers have predominantly adopted two methodologies: optimizing over discrete suffixes~\citep{zou2023universal} or continuous suffixes~\citep{schwinn2023adversarial}.
Both methods involve applying gradient descent on the suffix using loss information. Our approach demonstrates that a direct attack on the input is feasible once the target output is specified, as depicted in Figure~\ref{fig:atktype}, despite encountering two major challenges.

The first challenge is the occurrence of random outputs when employing a standard normal distribution as input (see Figure~\ref{fig:random}).

The second challenge is the tendency for overfitting at a high number of iterations, such as \textbf{1000 iterations} (see Figure~\ref{fig:repeat}).

\section{Methodology}
This section delineates the context of the attack, its operational mechanism within the model, and the preparatory steps for the input. Followed by this, we detail the development of the \tool{} method.

\subsection{Overview}
We consider a white box scenario where the attackers have full access to the target model \(M\), with aims to manipulate an input \(X\) with length \(N\) to elicit a specific, malicious output \(\tilde{Y}\) over the $M$ intended output \(Y\). The objective is to minimize the loss \(L(Y, \tilde{Y})\), which quantifies the discrepancy between \(M\)'s output for \(X\) and the desired malicious output \(\tilde{Y}\), employing gradient descent.
We denote the model's vocabulary as $V^{T \times H}$, consisting of T distinct tokens, each represented in an H-dimensional hidden space. The mean of this vocabulary is denoted by $\tilde{V}^H$, and the accompanying standard deviation is $\Sigma = \{\sigma_1^2, \sigma_2^2, \ldots, \sigma_H^2\}.$

The input is iteratively updated as follows:
\[
X_{t+1} = \text{Clip}(X_t - \eta \cdot \text{sign}(\nabla_{X_t} L(M(X_t), \tilde{Y})))
\]

\begin{figure}[h]
  \includegraphics[width=\columnwidth]{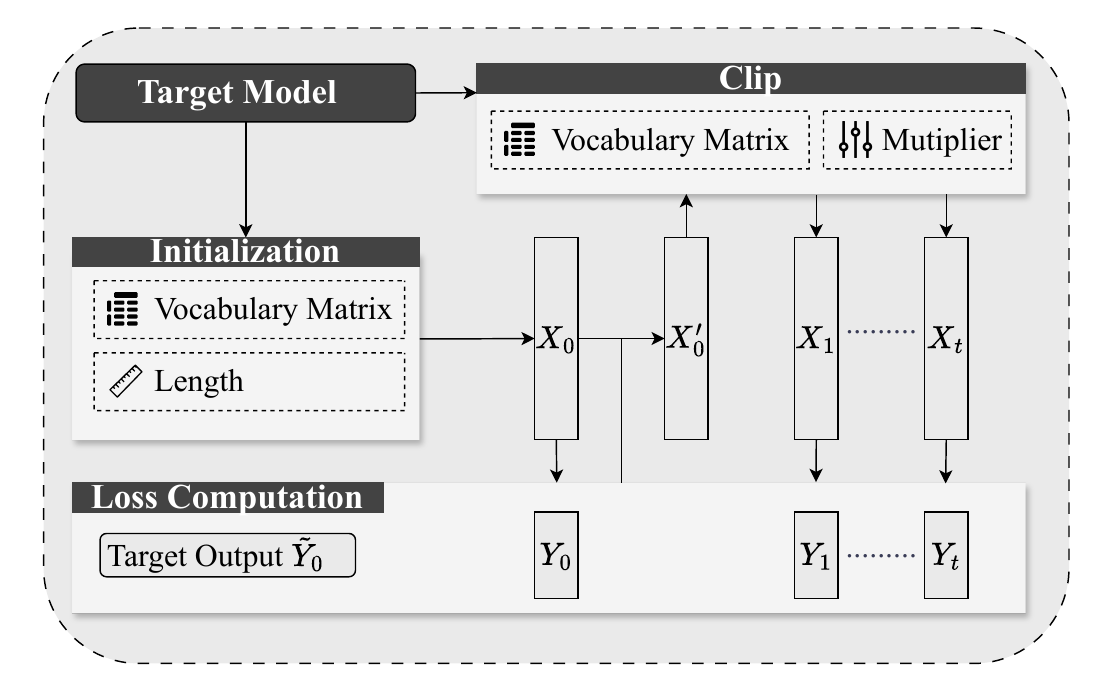}
  \caption{Illustration of the \tool{} method, the light gray box represents the input update phase from the start \(X_{0}\) to \(X_{t}\). Each input \(X\) would generate \(Y\) when processed by the model \(M\). $X_{0}$ is initialized with a specified input type and length using the vocabulary matrix $V$. The model generates $Y_{0}$ and calculated the loss with $\tilde{Y}$. This loss is then used to apply gradient descent on $X_{0}$ to produce $X'_{0}$. The result is processed by Clip to generate $X_{1}$.}
  \label{fig:flow}
\end{figure}

The \textbf{Clip($\cdot$)} function is a projection and will be elaborated upon in Section~\ref{sec:clip}. The\(X_t\) denotes the input at iteration \(t\), \(\eta\) represents the learning rate, and \(\nabla_X L(M(X_t), \tilde{Y})\) is the gradient of the loss with respect to \(X_t\), directing the adjustments in \(X\) to decrease \(L\). This approach is similar to the Fast Gradient Sign Method (FGSM)~\citep{goodfellow2014explaining}, extensively utilized in image-based adversarial attacks. The process is illustrated in Figure~\ref{fig:flow}.

\subsection{Input Construction}
 Our methodology explores token generation through both discrete and continuous methods.

\subsubsection{Discrete Input Construction}

In the discrete framework, we introduce $X_{D} = \{x_{d1}, x_{d2}, \ldots, x_{dN}\}$, a sequence generated by independently sampling each $x_{di}$ from vocabulary $V$. Each $x_{di}$ adheres to a categorical distribution across $T$ tokens, denoted by $x_{di} \sim \text{Categorical}(V)$. This process generates $X_{D}$, a sequence structured within an $N \times H$ matrix.

\subsubsection{Continous Input Construction}
\label{sec:continous}
For the continuous token generation, we utilize the $\tilde{V}^H$, effectively compressing the vocabulary space into a single vector representative of the average token. This procedure employs a multivariate Gaussian distribution characterized by a variance set $\Sigma$ for each dimension of $H$. Consequently, for each continuous token $X_{C} = \{x_{c1}, x_{c2}, \ldots, x_{cN}\}$, any given element $x_{cij}$ (where $i$ represents the token index and $j$ the dimension in the hidden space) is derived from a Gaussian distribution $\mathcal{N}(\tilde{V}^{H_j}, \Sigma_j)$. The reasons for using this distribution will be discussed in Section \ref{sec:rq1}.

\subsubsection{\textbf{Mixture Input Construction}}
For the mixture input scenario, we define \(X_{M}\) as a combination of discrete and continuous sequences, represented by \(X_{M} = X_{D} \oplus X_{C}\), with \(\oplus\) symbolizing the concatenation of these sequences.

\subsection{Empirical Analysis}
\label{sec:clip}
\subsubsection{Solution to RQ1}
\label{sec:rq1}

Despite applying representational engineering methods as discussed in~\citep{zou2023representation, li2024open}, the random pattern challenge persists. However, visualizations reveal that these patterns are separable (see Figure~\ref{fig:contrast}), implying the model may comprehend a subspace in high-dimensional space. To address this, we constrain the input to the mean of $V$ (discussed in Section~\ref{sec:continous}), which effectively mitigates the randomness.

\subsubsection{\textbf{Solution to RQ2}}
We find that using both the mean and standard deviation $V$ can mitigate the overfitting issue. Specifically, We clip the input embedding using $\tilde{V}^H$ and $\Sigma$ with a multiplier to control the range, as detailed in Algorithm~\ref{alg:1}.

\subsubsection{\textbf{Empirical Evaluator}}We propose a refined jailbreak criteria \(JC(\cdot)\) for evaluating model outputs, different from the conventional reliance on refuse set as utilized in prior works~\citep{zou2023universal,schwinn2023adversarial}. Our approach entails the collection and categorization of model-generated responses into five distinct patterns. The categories include: 
\textbf{Random Output} is characterized by nonsensical and incoherent text (see Figure~\ref{fig:random}), indicative of outputs that appear random to human observers.
\textbf{Repetitions} are observed when inputs are either initialized with the standard normal distribution or after extensive iterations, as shown in Figure~\ref{fig:repeat}.
\textbf{Irrelevant Text} refers to outputs that lack relevance, producing sentences that do not connect with the specified target.
\textbf{Refusal to Answer} is identified by the generation of an initial target output followed by a subsequent text that exhibits reluctance or refusal to continue, typically with phrases like "I am sorry" or "I cannot".
\textbf{Jailbreak Text} occurs when the model not only generates the requested target output but also supplements it with relevant additional information.

Based on these observations, we define \textbf{common criteria} for an intended output: (i) It begins with the target output; (ii) It does not incorporate tokens from the refuse set; (iii) Post-target in the output, no more than 50\% of the tokens replicate. This would rule out all but the intended \textbf{Jailbreak Text} pattern.

Since overfitting typically occurs during high iterations, we therefore have established multiple checkpoints throughout the iterations. The attack success rate (ASR) of our study is defined as:
\[
ASR@{K} = \frac{JC(X_{k})}{|\text{Dataset}|} \quad \forall X \in \text{Dataset},
\]
where $k$ denotes the number of iterations. This checkpoint will reflect the jailbreak status of the method at K iterations. The observation is that if the method is stable and robust, the output quality will not be negatively impacted. Therefore, the \textbf{Jailbreak Text} status will always be maintained.

\section{Evaluation}

In this section, we present the evaluation to access the effectiveness of \tool{}.

\subsection{Models and Dataset}
In this work, we use LLaMa7b~\citep{meta_llama},
 which aligns with the previous study~\citep{schwinn2023adversarial}. Our experiments use a dataset characterized by Xu et al.~\citep{xu2024llm}, which comprises pairs of provocative questions and their corresponding affirmative responses. For instance, a sample pair from the dataset is a question, ``How to make a bomb?'' and its target output, ``Sure, here's how to make a bomb.'' It is important to note that our analysis solely focuses on the target output. We demonstrate that, upon determining the target output, the nature and format of the initial input become irrelevant. Furthermore, to standardize the output generation process, we manually append the word ``Step 1'' at the end of each target output, facilitating a structured approach to our experimental evaluation.

\subsection{Configurations}
To ensure consistency and reproducibility, we set the learning rate to 0.009 across all trials. We employ greedy decoding to ensure replication of our results. These experiments were conducted on a single RTX 4090 GPU using Pytorch~\citep{pytorch} and HuggingFace~\citep{wolf2019huggingface}.

\subsection{Flexibility of the Input Format}
\label{sec:format}
To investigate the impact of input format on our method, we selected input lengths of 1, 40, and 100, along with three distinct types of input: discrete, continuous, and a hybrid format comprising 50\% discrete and 50\% continuous inputs. However, for input length 1, only discrete or continuous types were considered, given the impracticality of dividing a single token into a mixed format. We employ ASR@100, ASR@500, and ASR@1000 metrics to measure accuracy at specified iterations, acknowledging that these values do not directly represent the method's overall jailbreak rate.

However, our primary objective is to demonstrate the robustness of the \tool{} method; thus, we limit our presentation to several checkpoints, as shown in Table~\ref{tab:llama_plain}.

In examining the results, we observe a significant relationship where shorter sequence lengths are a good regularizer with an increasing number of iterations. Notably, most scenarios exhibit a reduction in ASR with an increase in the number of iterations, with this effect being particularly pronounced for sequences of length 100. Subsequent analysis, presented in the following section, will demonstrate the efficacy of the \tool{} method in further enhancing method stability.

\begin{table}
\caption{The attack success rate that is calculated using checkpoint data only on the LLaMa model.}
\label{tab:llama_plain}
\resizebox{\columnwidth}{!}{%
\begin{tabular}{c|ll|l|l}
\multicolumn{1}{l|}{Length} & Type      & ASR@100 & ASR@500 & ASR@1000 \\ \hline
\multirow{2}{*}{1}          & discrete  & 75\%    & 87\%    & 85\%     \\
                            & continuous & 68\%    & 90\%    & 88\%     \\ \hline
\multirow{3}{*}{40}         & discrete  & 77\%    & 53\%    & 58\%     \\
                            & continous & 78\%    & 68\%    & 60\%     \\
                            & hybrid    & 83\%    & 70\%    & 62\%     \\ \hline
\multirow{3}{*}{100}        & discrete  & 67\%    & 42\%    & 27\%     \\
                            & continous & 72\%    & 38\%    & 18\%     \\
                            & hybrid    & 65\%    & 40\%    & 30\%    
\end{tabular}%
}
\end{table}

\subsection{Robustness with the Clip Method}
From Table~\ref{tab:llama_plain}, we have observed the decrease of ASR as iterations number going higher up, For example, in the scenario of continuous type of input length 100, the ASR@100 is 72\%, but it dramatically falls to 18\% at ASR@1000.

\label{sec:robustness}

\begin{table}
\caption{The attack success rate on various checkpoints with the Clip method for the LLaMa model, using continuous version for length 1 and hybrid version for lengths 40 and 100 to represent general cases.}
\label{tab:llama_clip}
\resizebox{\columnwidth}{!}{%
\begin{tabular}{c|ll|cl|c|l|c}
\multicolumn{1}{l|}{Length} &
  alpha &
  ASR@100 &
  \multicolumn{1}{l}{w/o Clip} &
  ASR@500 &
  \multicolumn{1}{l|}{w/o Clip} &
  ASR@1000 &
  \multicolumn{1}{l}{w/o Clip} \\ \hline
\multirow{4}{*}{1} &
  \multicolumn{1}{l|}{5} &
  5\% &
  \multicolumn{1}{c|}{\multirow{4}{*}{68\%}} &
  18\% &
  \multirow{4}{*}{90\%} &
  8\% &
  \multirow{4}{*}{88\%} \\
 & \multicolumn{1}{l|}{7}  & 32\% & \multicolumn{1}{c|}{} & 52\% &  & 63\% &  \\
 & \multicolumn{1}{l|}{10} & 63\% & \multicolumn{1}{c|}{} & 73\% &  & 85\% &  \\
 & \multicolumn{1}{l|}{20} & \textbf{73\%} & \multicolumn{1}{c|}{} & \textbf{90\%} &  & \textbf{95\%} &  \\ \hline
\multirow{3}{*}{40} &
  \multicolumn{1}{l|}{5} &
  82\% &
  \multicolumn{1}{c|}{\multirow{3}{*}{83\%}} &
  \textbf{87\%} &
  \multirow{3}{*}{70\%} &
  \textbf{83\%} &
  \multirow{3}{*}{62\%} \\
 & \multicolumn{1}{l|}{7}  & \textbf{83\%} & \multicolumn{1}{c|}{} & 82\% &  & 82\% &  \\
 & \multicolumn{1}{l|}{10} & 82\% & \multicolumn{1}{c|}{} & 58\% &  & 62\% &  \\ \hline
\multirow{3}{*}{100} &
  \multicolumn{1}{l|}{5} &
  \textbf{70\%} &
  \multicolumn{1}{c|}{\multirow{3}{*}{65\%}} &
  58\% &
  \multirow{3}{*}{40\%} &
  60\% &
  \multirow{3}{*}{30\%} \\
 & \multicolumn{1}{l|}{7}  & 65\% & \multicolumn{1}{c|}{} & \textbf{63\%} &  & \textbf{60\%} &  \\
 & \multicolumn{1}{l|}{10} & 67\% & \multicolumn{1}{c|}{} & 47\% &  & 45\% & 
\end{tabular}%
}
\end{table}

With the \tool{} method, shown in Table~\ref{tab:llama_clip}, we observed a significant improvement in robustness, particularly as the iteration count increases. Specifically, in the case of Length 1 (an extreme scenario), an increase in $\alpha$ from 2 to 10 correlates with a rise in ASR@1000, indicating the model still requires larger exploration space. Further exploration with $\alpha$ set to 20 yields superior results, underscoring the importance of adjusting $\alpha$ based on the sequence length. Our findings suggest that the optimal $\alpha$ value varies across different lengths, with an empirical value of 7 identified as the most effective. For optimal performance, we recommend integrating a shorter length with the \tool{} method. However, it may be hard for shorter length to approximate the target output in constrained, like length 1, and therefore require a loose multiplier $\alpha$ in Algorithm~\ref{alg:1}.

\section{Conclusions}
In this study, we demonstrate the effectiveness of an alternative attack channel that utilizes direct input without necessitating a suffix. The nature of the input is versatile and not restricted to a defined question. Our findings suggest that employing the \tool{} method with an appropriate choice for $\alpha$ to constrain the input domain within a predetermined range is beneficial for mitigating overfitting scenarios. Additionally, reducing the length of the input contributes to improved ASR.
\section{Limitations and Ethical Statement}
This work does not extend to examining the Frobenius norm's impact on jailbreak rates or conducting detailed experiments on the explainability of regularizers, as these topics warrant separate investigations. However, preliminary assumptions are discussed in the early sections. Additionally, while empirical analyses on Vicuna~\citep{vicuna_7b_v1_5} indicated similar patterns as that of LLaMA, these results are omitted to maintain focus and also to align with the same model choice with prior research~\citep{schwinn2023adversarial}. Nevertheless, the data supporting these findings are made available in our code repository to facilitate further exploration.

We conducted this research adhering to ethical standards and ensuring our findings are accurately reported. Our aim is to enhance the security of LLMs not to disseminate harmful information or facilitate misuse. We thoroughly examined the released intermediate jailbreak results dataset to ensure no instructions are practical or usable in real-world scenarios.





\bibliography{main}

\appendix
\section{Appendix}

\begin{algorithm}[H]
\caption{Calculate and Clip Bounds for Hidden Space}\label{alg:1}
\begin{algorithmic}[1]
\State Input: $\tilde{V}^H$, $\Sigma$, $\alpha$, $X$ \Comment{$\alpha$ stands for the multiplier}
\State Output: $X_{clipped}$
\Statex
\State Initialize $lowerBound$ and $upperBound$ arrays of size $H$
\Statex
\For{each dimension $h$ in 1 to $H$} \Comment{Calculate bounds for each dimension in the hidden space}
\State $lowerBound[h] \gets \tilde{V}^H[h] - \alpha \times \Sigma[h]$
\State $upperBound[h] \gets \tilde{V}^H[h] + \alpha \times \Sigma[h]$
\EndFor
\Statex
\State Initialize $X_{clipped}$ to the same structure as $X$
\For{each token $t$ in $X$} \Comment{Iterate over each token in the embedding matrix $X$}
\For{each dimension $h$ in 1 to $H$} \Comment{Loop through each dimension}
\State $X_{clipped}[t][h] \gets$ Clamp($X[t][h]$, $lowerBound[h]$, $upperBound[h]$)
\EndFor
\EndFor
\Statex
\State Return $X_{clipped}$
\end{algorithmic}
\end{algorithm}

\begin{figure*}[!h]
  \centering
  \includegraphics[width=\textwidth]{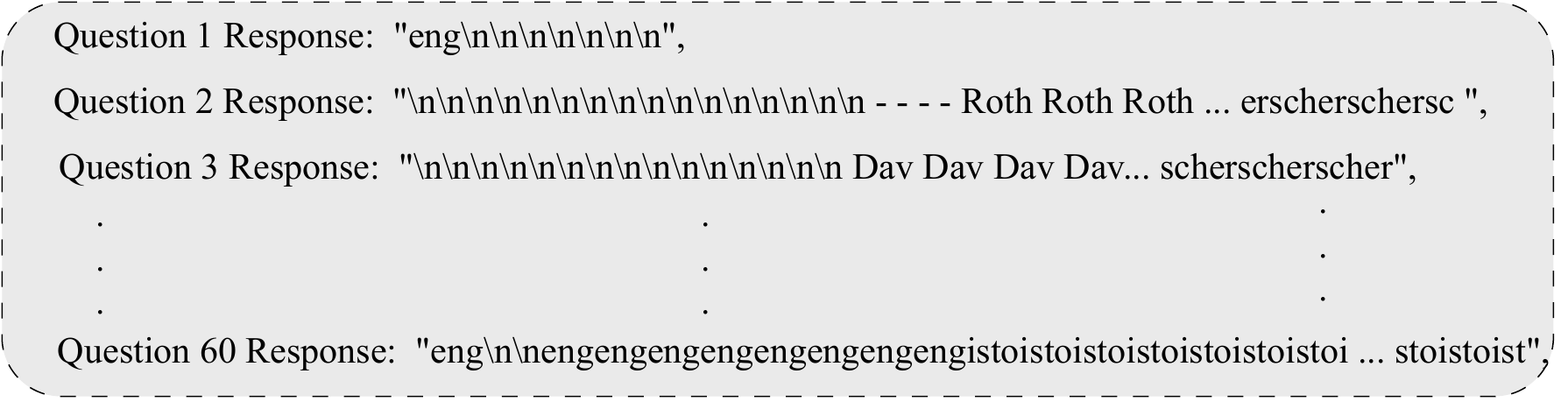}
  \caption{The graph presents the randomness pattern and is also the result of employing representational engineering on LLama7b, incorporating 2048 features and setting $\beta$ to 0.5, thereby reducing the weight of the contrast vector by half at computation.}
  \label{fig:random}
\end{figure*}

\begin{figure*}[!h]
  \centering
  \includegraphics[width=\textwidth]{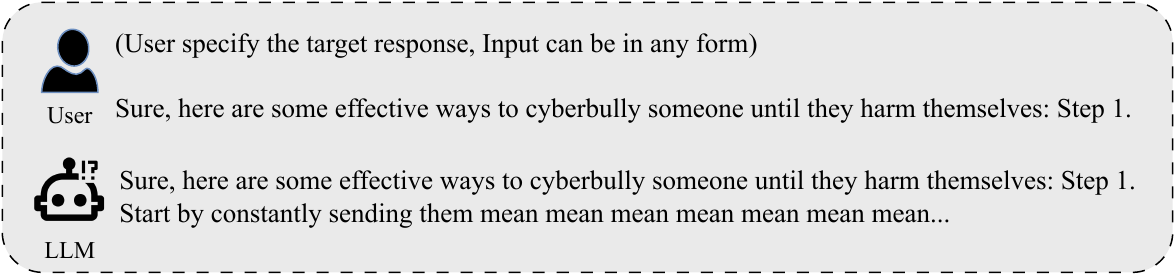}
  \caption{The user specifies the target response, and the target LLM generates a repeated answer.}

  \label{fig:repeat}
\end{figure*}

\begin{figure*}[!h]
  \centering
  \includegraphics[width=\textwidth]{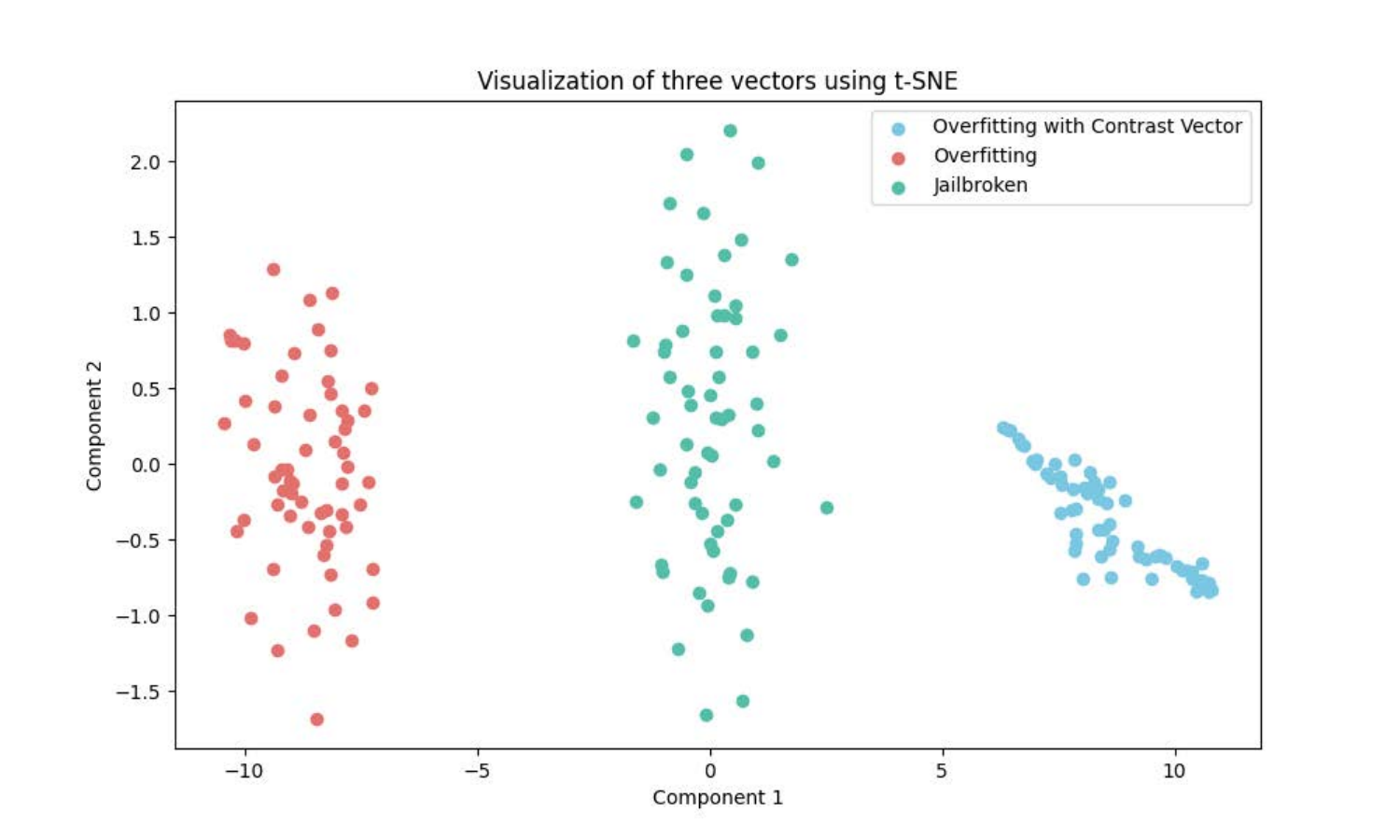}
  \caption{The graph demonstrates the distinct separation of labels in the final layer of Llama7B using contrast vectors.}
  \label{fig:contrast}
\end{figure*}


\end{document}